\documentclass{aastex63}
\usepackage[caption=false]{subfig}
\usepackage{graphicx}
\usepackage{multirow}
\usepackage{graphicx}
\usepackage{booktabs}
\usepackage{amsmath,amssymb}
\usepackage[T1]{fontenc}
\usepackage{epstopdf}
\graphicspath{{./}{figures/}}

\begin{document}

\title{X-ray and GeV-$\gamma$-ray emission property of TeV Compact Symmetric Object PKS 1413+135 and Implication for Episodic Jet Activity}

\correspondingauthor{Jin Zhang}
\email{j.zhang@bit.edu.cn}

\author[0000-0002-4789-7703]{Ying-Ying Gan}
\affiliation{School of Physics, Beijing Institute of Technology, Beijing 100081, People's Republic of China; j.zhang@bit.edu.cn}

\author[0000-0003-3554-2996]{Jin Zhang\dag}
\affiliation{School of Physics, Beijing Institute of Technology, Beijing 100081, People's Republic of China; j.zhang@bit.edu.cn}

\author[0000-0002-9728-1552]{Su Yao}
\affiliation{Max-Planck-Institute f\"{u}r Radioastronomie, Auf dem H\"{u}gel 69, 53121 Bonn, Germany}

\author[0000-0001-6863-5369]{Hai-Ming Zhang}
\affiliation{School of Astronomy and Space Science, Nanjing University, Nanjing 210023, People's Republic of China}

\author[0000-0002-6316-1616]{Yun-Feng Liang}
\affiliation{Guangxi Key Laboratory for Relativistic Astrophysics, School of Physical Science and Technology, Guangxi University, Nanning 530004, People's Republic of China}

\author[0000-0002-7044-733X]{En-Wei Liang}
\affiliation{Guangxi Key Laboratory for Relativistic Astrophysics, School of Physical Science and Technology, Guangxi University, Nanning 530004, People's Republic of China}

\begin{abstract}
PKS 1413+135, a compact symmetric object (CSO) with a two-side pc-scale structure in its miniature radio morphology, is spatially associated with 4FGL J1416.1+1320 and recently detected with MAGIC telescopes. We comprehensively analyze its X-ray and GeV $\gamma$-ray observation data for revealing its high energy radiation physics. It is found that the source is in a low-flux stage before MJD 58500 and experiences violent outbursts after MJD 58500 in the GeV band. The flux at 10 GeV varies $\sim 3$ orders of magnitude, and the GeV-flux variation is accompanied by the clear spectral variation, which is characterized as a soft log-parabola spectrum in the low-flux state and a hard power-law spectrum in the bright flares. The variability amplitude of X-rays is lower than that of $\gamma$-rays, and no correlation of variability between $\gamma$-rays and X-rays is observed. Fitting the broadband spectral energy distribution during a GeV outburst with a multi-zone leptonic model, we show that the GeV $\gamma$-rays are attributed to the external Compton process while the X-rays are hybrid of several components. The predicted TeV $\gamma$-ray flux during the GeV outburst is consistent with the detection of MAGIC telescopes. These results, together with its CSO radio morphology, imply that PKS 1413+135 has the episodic nuclear jet activities. The weak $\gamma$-ray emission before MJD 58500 may be from its sub-pc-/pc-scale jet component powered by previous activities, and the violent outbursts with short-timescale variability after MJD 58500 could be attributed to the recently restarted jet activity.
 \end{abstract}


\section{Introduction}

Compact symmetric objects (CSOs), a sub-class of active galactic nuclei (AGNs), are defined as those with symmetric twin radio jet structure on both sides of their nuclei (e.g., \citealp{1996ApJ...460..612R}). They are thought to be a class of misaligned AGNs (\citealp{1994ApJ...432L..87W, 1996ApJ...460..612R}). Generally, low polarization, low radio variability, low core luminosity, hosted in elliptical galaxies, and lack of the optically thick component at short wavelengths are presented in CSOs (\citealp{1994ApJ...432L..87W, 1996ApJ...460..612R}). CSOs are characterized by an overall size being less than 1 kpc. The small size of CSOs may result from the frustrated jet due to the dense interstellar medium (\citealp{1991ApJ...380...66O, 1994A&A...292..392C, 1998A&A...329..845C}). The ages estimated by the advance speed, the lobe supply timescale, and the synchrotron-loss timescale for CSOs are consistent with each other within the uncertainties, indicating that CSOs are young with typical ages of 3$\times10^3-10^4$ yr, almost certainly $\ll 10^5$ yr (\citealp{1996ApJ...460..612R}). The non-difference between these compact radio sources and larger radio-loud objects at the mid- and far-infrared emission also supports that CSOs are young (\citealp{1994ApJ...428...65H}). Thus, CSOs are an important fraction of radio-loud AGNs for understanding the formation and evolution of powerful jets in extragalactic radio sources.

Gamma-ray emission is a critical probe to study the AGN jets. As a new population of $\gamma$-ray sources, six CSOs have been detected with the \emph{Fermi}-LAT in the GeV band, i.e., PMN J1603--4904 (\citealp{2014A&A...562A...4M, 2015A&A...574A.117M}), PKS 1718--649 (\citealp{2016ApJ...821L..31M}), NGC 3894 (\citealp{2020A&A...635A.185P}), TXS 0128+554 (\citealp{2020ApJ...899..141L}), CTD 135 (\citealp{2021RAA....21..201G}), and PKS 1413+135 (\citealp{2021MNRAS.507.4564P}). It is debated that whether the $\gamma$-rays of CSOs are from the pc-scale lobes or from the core jet region. Interestingly, the $\gamma$-rays in some CSOs may indicate a characteristic of recently restarted jet activity (\citealp{2020ApJ...899..141L, 2021RAA....21..201G}). Recently, PKS 1413+135 was announced to be detected by the Major Atmospheric Gamma-ray Imaging Cherenkov (MAGIC) telescopes on January 12, 2022 (MJD 59591, \citealp{2022ATel15161....1B}). It is the first CSO detected at the very-high-energy (VHE; $E > 100 \rm \,GeV$) band. It would be a good sample to study the $\gamma$-ray emission of young radio sources.

The high-resolution observations with the very large array (VLA), US very long baseline interferometry (VLBI) Network, and very long baseline array (VLBA) show no kpc-scale extended structure is presented for PKS 1413+135, and there is a mini-triple structure with a counter-jet on pc-scale (\citealp{1994ApJ...424L..69P}). Using the high-dynamic range VLBA maps at 3.6, 6, 13, and 18 cm, \cite{1996AJ....111.1839P} revealed a complex, two-side pc-scale structure for PKS 1413+135 and no evidence of superluminal motion was obtained, and thus they suggested that PKS 1413+135 likely is a young radio source with an age $\leq10^4$ yr. It is intriguing that the multi-band optical images of PKS 1413+135 show a surface brightness profile that can be well fitted by an exponential disk, suggesting that PKS 1413+135 is hosted in a spiral galaxy (\citealp{1991MNRAS.249..742M}). The Hubble Space Telescope (HST) observation further reveals that the galaxy contains a previously unresolved dust lane and very likely is an early-type spiral galaxy viewed edge-on (\citealp{1994MNRAS.268..681M}). In addition, an extremely large column density is needed in the analysis of \emph{Einstein} X-ray data (\citealp{1992ApJ...400L..17S}). The enormous extinction, but the lack of the thermal IR emission and strong narrow emission lines, suggests that PKS 1413+135 would not be hosted in this spiral galaxy and be a background source behind the spiral galaxy (\citealp{1992ApJ...400L..17S}). The redshift of z=0.247 for PKS 1413+135 is derived from the redshift HI absorption (\citealp{1992ApJ...400L..13C}), which should be associated with the spiral galaxy. The U-shaped symmetric achromatic variability from 15 to 234 GHz of PKS 1413+135 was proposed to be caused by gravitational lensing of the foreground spiral galaxy (\citealp{2017ApJ...845...89V}). However, no sign of image multiplicity or distortion due to the gravitational lensing is found, in either optical or radio bands (\citealp{1992ApJ...400L..17S, 1996AJ....111.1839P, 2017ApJ...845...89V, 2021ApJ...907...61R}). It was thus argued that PKS 1413+135 should be located at a redshift range of $0.247<z<0.5$ (\citealp{2017ApJ...845...89V, 2021ApJ...907...61R}). Due to the strong obscuring and contamination by the foreground, the high energy photons, especially $\gamma$-rays in the GeV--TeV bands, would be powerful probes for studying the nature of the source.

For revealing the radiation properties of PKS 1413+135, we analyze its data in the X-ray and $\gamma$-ray bands observed with the \emph{Chandra}, \emph{XMM-Newton}, \emph{Swift}-XRT, and \emph{Fermi}-LAT over the past $\sim$ 13.5 yr. Description of the data reduction is presented in Section 2. We investigate its temporal and spectral variations in both X-ray and $\gamma$-ray bands (Section 3 and Section 4). Together with the data in the low-energy bands, we compile its broadband spectral energy distributions (SEDs) and study its $\gamma$-ray emission via the SED modeling in Section 5. Discussion on the $\gamma$-ray origin of PKS 1413+135 and its AGN type is given in Section 6. A summary is presented in Section 7. Throughout, if not otherwise specified, the results are derived on the basis of $z=0.247$. $H_0=71$ km s$^{-1}$ Mpc$^{-1}$, $\Omega_{\rm m}=0.27$, and $\Omega_{\Lambda}=0.73$ are adopted in this paper.

\section{Observations and Data Reduction}

\subsection{GeV $\gamma$-Ray Observations with Fermi-LAT}

It was reported that 4FGL J1416.1+1320 is spatially coincident with PKS 1413+135 in the \emph{Fermi}-LAT 12-year Source Catalog (4FGL-DR3, \citealp{2022ApJS..260...53A}). We download the Pass 8 data from the Fermi Science Support Center (FSSC). The data are selected within a $15^{\circ}$ region of interest (ROI) centered on the radio position of PKS 1413+135 (R.A. = $213.995\degr$, Decl. = $13.340\degr$; \citealp{2015AJ....150...58F}). The temporal coverage of the data is from 2008 August 04 to 2022 March 22 (MJD 54682--59660) of about 13.5 yr. We perform a binned likelihood analysis for the $\gamma$-rays of 4FGL J1416.1+1320 using the publicly available software \textit{Fermitools} (ver. 2.0.8). Only the $\gamma$-ray photons in the energy range of 0.1--300 GeV and satisfying the standard data quality selection criteria ``$(DATA\_QUAL > 0)  \&\& (LAT\_CONFIG == 1)$'' are considered in our analysis. A zenith angle cut of $90^{\circ}$ is set to avoid the $\gamma$-ray contamination causing by the Earth limb. We bin the data with a pixel size of $0.2^{\circ}$ in space and 25 logarithmical energy bins. The background models include all $\gamma$-ray sources listed in the 4FGL-DR3 Catalog and the Galactic diffuse component (gll\_iem\_v07.fits) as well as the isotropic emission (iso\_P8R3\_SOURCE\_V3\_v1.txt). The P8R3\_SOURCE\_V3 set of instrument response functions (IRFs) is used.

The spectrum model of 4FGL J1416.1+1320 reported in the 4FGL-DR3 is a log-parabola function (\citealp{2022ApJS..260...53A}), i.e.,
\begin{equation}
\label{LAT_spec}
\frac{dN}{dE} =  N_{0}(\frac{E}{E_{\rm b}})^{-(\Gamma_{\gamma}+\beta{\log}(\frac{E}{E_{\rm b}}))},
\end{equation}
where the decimal logarithm is used, $E_{\rm b}$ is the scale parameter of photon energy, $\Gamma_{\gamma}$ is the photon spectral index, and $\beta$ is the curvature parameter (\citealp{2004A&A...413..489M}). So, the model for fitting the spectrum of 4FGL J1416.1+1320 is in priority selected as a log-parabola function in our analysis. If $\beta$ is poorly constrained, we set $\beta=0$ and the log-parabola function turns into a single power-law function (i.e., $dN/dE = N_{0}(E/E_{\rm b})^{-\Gamma_{\gamma}}$). The spectral parameters of all sources lying within $8^{\circ}$ are left free, whereas the parameters of those sources lying beyond $8^{\circ}$ are fixed to their 4FGL-DR3 values. Also, the normalization parameters of the standard Galactic and isotropic background templates are set free in the likelihood fit. The significance of the $\gamma$-ray detection is quantified by adopting the maximum likelihood test statistic (TS), which is defined as TS$= 2 (\ln\mathcal{L}_{1}-\ln\mathcal{L}_{0})$, where $\mathcal{L}_{1}$ and $\mathcal{L}_{0}$ are maximum likelihood values for the models with and without the target source, respectively. The analysis results of the \emph{Fermi}-LAT observations for PKS 1413+135 are given in Table \ref{table1}.

\subsection{X-ray observations with Swift-XRT, Chandra, and XMM-Newton}

PKS 1413+135 was observed by the \emph{Swift}-XRT, \emph{Chandra}, and \emph{XMM-Newton} in several epochs, as listed in Table \ref{table2}. It was observed by the \emph{Chandra} Advanced CCD Imaging Spectrometer S-array (ACIS-S) as a general observer target for an exposure of 20\,ks on March 20, 2007. It was also observed several times by \emph{Chandra} again during 2019 and 2020 as a Director's Discretionary Time target for exposures of 4\,ks. As a significant foreground absorption of soft X-rays was found in PKS 1413+135 (\citealp{1992ApJ...400L..17S}), firstly we attempt to get the equivalent neutral hydrogen column density of this absorption using the observation in 2007. The data are reduced using {\sc ciao} (version 4.12) and {\sc caldb} (version 4.9.2.1). The level-2 event file is created following the standard procedure. We extract source photons from a circle centred on the radio position of PKS 1413+135 with radius of 6 arcsec. The background is determined in an annulus region with inner and outer radius of 7 and 14 arcsec, respectively. There are $\sim600$ net photon counts. The spectrum is grouped to have at least 25 counts per bin and the $\chi^{2}$ minimization technique is adopted for spectral analysis. The spectrum is fitted by a single power-law absorbed by two absorption components, one is an absorption at $z=0$ with neutral hydrogen column density fixed at Galactic value $N^{\rm H}_{\rm gal}=1.56\times10^{20}$cm$^{-2}$ (\citealp{2016A&A...594A.116H}), the other one is an extragalactic foreground absorption $N^{\rm H}_{\rm int}$ at redshift of $z=0.247$ with column density set free. The model gives a well fit to the spectrum and an extragalactic foreground absorption column density of $N^{\rm H}_{\rm int}=3.6^{+0.6}_{-0.5}\times10^{22}$cm$^{-2}$, which is consistent with the results in \cite{1992ApJ...400L..17S} and \cite{2002AJ....124.2401P}. We generate the \emph{Chandra} spectra at other epochs following the procedures mentioned above. The single power-law with the Galactic ($N^{\rm H}_{\rm gal}$) and extragalactic ($N^{\rm H}_{\rm int}$) absorption is used to fit the spectra. As there are only no more than 150 net photon counts for each spectrum, $C$-statistic minimization is adopted for evaluating the goodness of the fits and $N^{\rm H}_{\rm int}$ is fixed at $3.6\times10^{22}$cm$^{-2}$ during the fitting.

The public data of pointing observations on PKS~1413+135 by the \emph{XMM-Newton} are reduced with the XMM--Newton {\sc science analysis system} (version 18) following standard procedures. We generate the spectra using the data from PN CCD arrays since PN has a larger effective area. The source events are extracted from a circle of 32 arcsec radius centred at source position while the background events are extracted from a circle of the same radius in a source-free region nearby. The spectrum is binned to contain at least 25 counts per bin required for $\chi^{2}$ analysis. Again, the single power-law with a Galactic and an extragalactic absorption is used to fit the spectra. The absorption is also fixed as mentioned above during the fitting.

There are 21 observations by the Neil Gehrels \emph{Swift} observatory for PKS 1413+135 from 2007 to 2022. The X-ray telescope (XRT) onboard the \emph{Swift} satellite was operating in the photon counting mode with exposure times of 0.1--10\,ks. We collect the XRT data from the \emph{Swift} archive and reproduce the clean events using the {\it xrtpipeline} task. The source photons are extracted from a circle with radius of 50 arcsec, while the backgrounds are determined in an annulus with inner and outer radius of 60 and 105 arcsec, respectively. There remain 16 observations after excluding those without significant detection for this source. As the statistics are too low for each spectrum, $C$-statistic minimization is adopted to evaluate the goodness of the fits. $N^{\rm H}_{\rm gal}=1.56\times10^{20}$cm$^{-2}$ and $N^{\rm H}_{\rm int}=3.6\times10^{22}$cm$^{-2}$ are all fixed, and the photon index is also fixed at $\Gamma_{\rm X}=1.9$ (the average value in Figure \ref{spectra}(b)) during the fitting.

\section{Temporal Analysis}

\begin{figure}
 \centering
   \includegraphics[angle=0,scale=0.7]{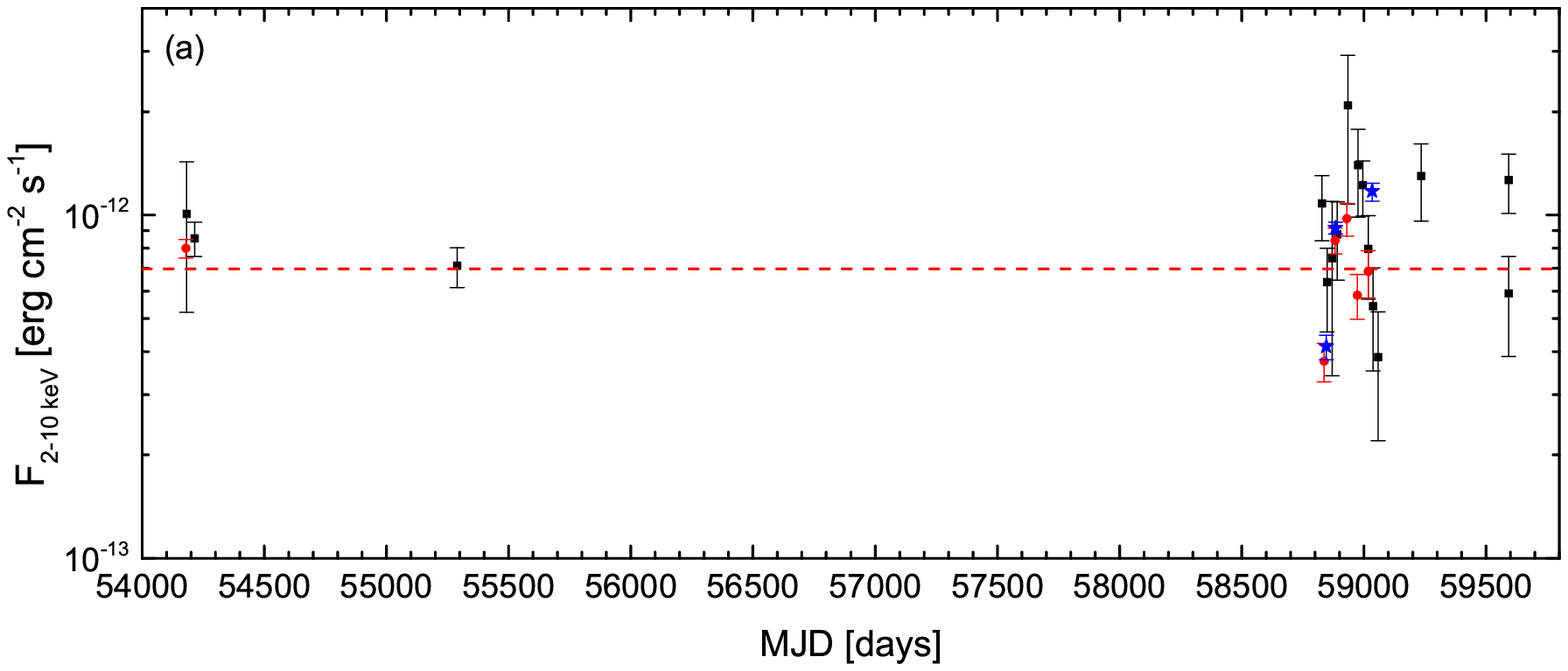}
   \includegraphics[angle=0,scale=0.7]{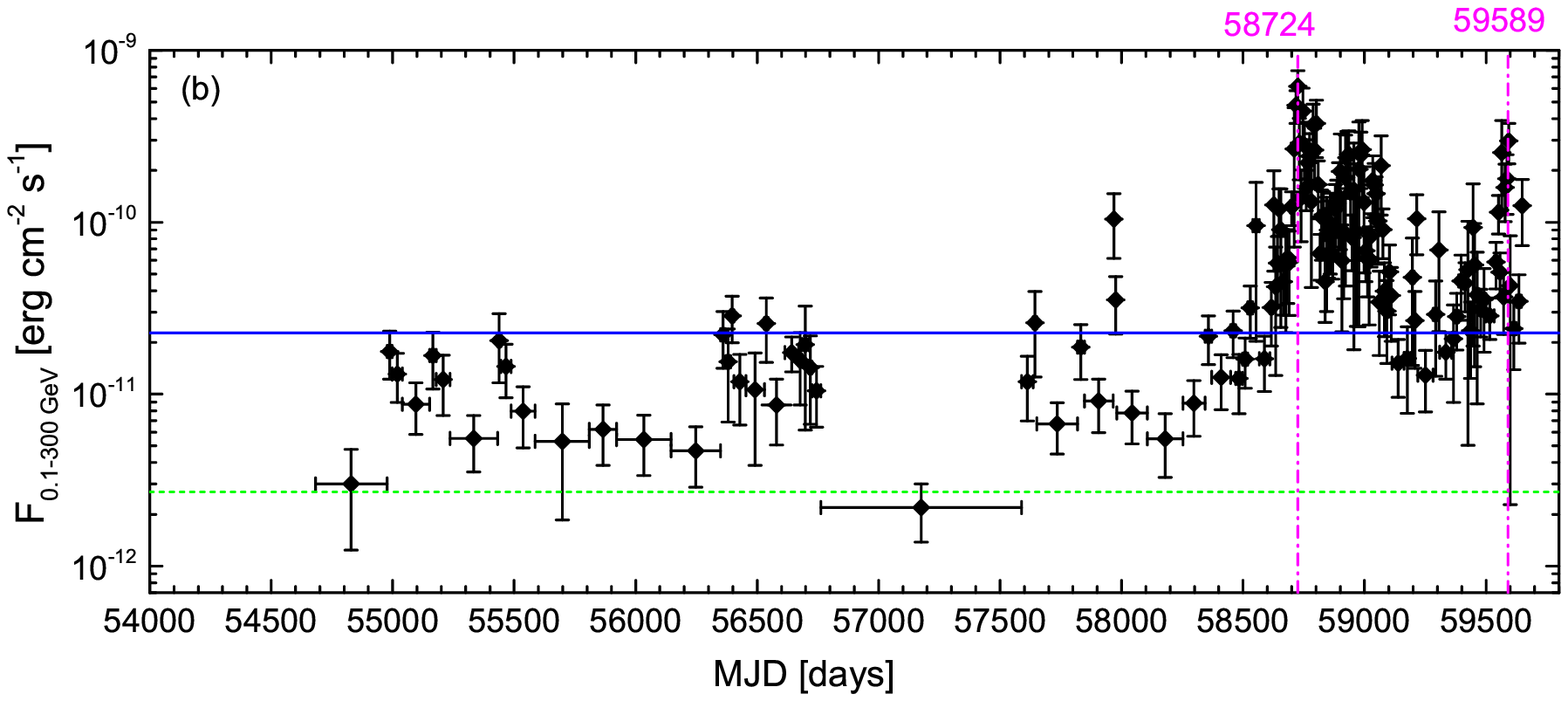}
   \includegraphics[angle=0,scale=0.7]{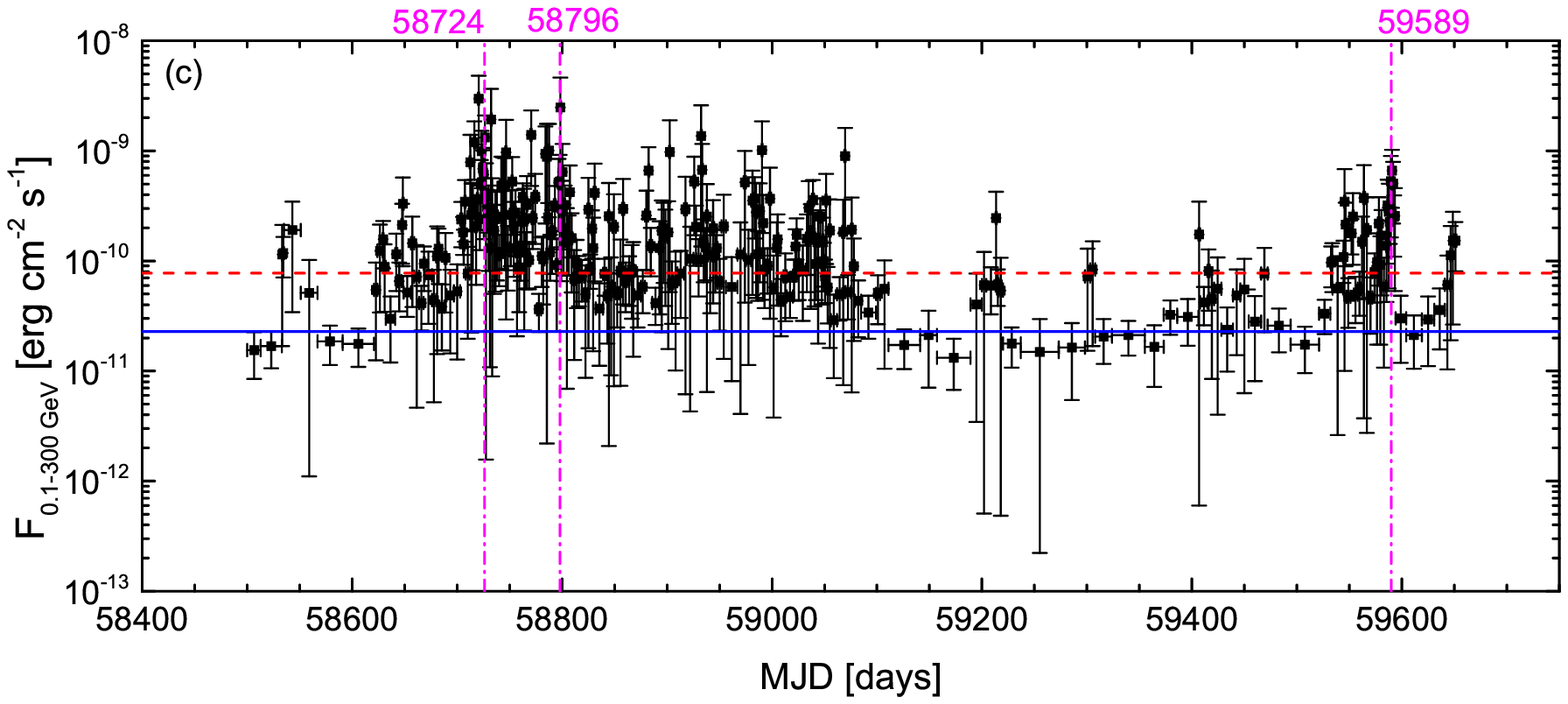}
\caption{Light curves of PKS 1413+135 in X-ray and $\gamma$-ray bands. \emph{Panel-(a)}: the X-ray light curve obtained by the observations of \emph{Swift}-XRT (black squares), \emph{Chandra} (red circles), and \emph{XMM-Newton} (blue stars), where the horizontal red dashed line represents the mean flux of all the data points derived with Equation (2). \emph{Panel-(b)}: the $\sim$13.5-year $\gamma$-ray light curve derived with an adaptive-binning method based on a criterion of TS$\geq9$ for each time bin, where the minimum time-bin step is 7 days. The horizontal blue solid line indicates the 13.5-year average flux of $2.28\times10^{-11}$ erg cm$^{-2}$ s$^{-1}$. The horizontal green short-dashed line is the sensitivity of \emph{Fermi}-LAT, which is obtained and derived from the link of https://www.slac.stanford.edu/exp/glast/groups/canda/lat\_Performance.htm. \emph{Panel-(c)}: the $\gamma$-ray light curve after MJD 58500 also obtained with the adaptive-binning method and the criterion of TS$\geq9$ for each time bin, where the minimum time-bin step is one day. The horizontal red dashed line indicates the average flux of $7.74\times10^{-11}$ erg cm$^{-2}$ s$^{-1}$ while the horizontal blue solid line is the same as in \emph{Panel-(b)}.}\label{LC}
\end{figure}

We perform a likelihood fit for PKS 1413+135 with the $\sim$13.5-year \emph{Fermi}-LAT data. PKS 1413+135 is bright in the GeV $\gamma$-ray band, with an average flux of $F_{0.1-300\rm\,GeV}=(2.28\pm0.13)\times10^{-11}$ erg cm$^{-2}$ s$^{-1}$ in the whole interval. Variability is very common for AGNs and its magnitude and timescale are useful to study the property of emission region. For estimating the variability of PKS 1413+135 at the GeV band, we follow the definition in 2FGL (\citealp{2012ApJS..199...31N}) and derive the variability index TS$_{\rm var}$ (see also \citealp{2020ApJS..247...33A}). We split the full $\sim$13.5-year interval into N=13 intervals of about one year each. The source is considered to be significantly variable if TS$_{\rm var}$ exceeds 53.9, where TS$_{\rm var}=53.9$ corresponds to the $5\sigma$ confidence level in a $\chi^2_{N-1}$(TS$_{\rm var}$) distribution with 12 degrees of freedom. The TS$_{\rm var}$ value of PKS 1413+135 is 2074 and far beyond $5\sigma$ confidence level, meaning that its $\gamma$-rays are significantly variable. We extract the $\gamma$-ray light curve of PKS 1413+135 using an adaptive-binning method based on a criterion of TS$\geq9$ for each time bin, where the minimum bin size is set to be seven days, as shown in Figure \ref{LC}(b). We note that before MJD 58500, the $\gamma$-ray emission is in a relatively low state, where the $\gamma$-ray fluxes are lower than the average flux level for almost all time bins and even lower than the average flux by almost an order of magnitude in some time bins. The $\gamma$-ray flux begins to increase from a very low state since $\sim$ MJD 58200 to the average flux level around MJD 58500. After MJD 58500 the source seems to be in the high state with fluxes higher than the average flux, and almost all the time bins only need the minimum bin size of seven days to meet TS$>$9. Therefore, the whole $\gamma$-ray light curve of PKS 1413+135 seems to present two totally different phases with a boundary of MJD 58500, i.e., a low-flux stage and a outburst stage.

In order to further investigate the short timescale variability of this source, we reanalyze the GeV observation data after MJD 58500 and still extract the light curve using the adaptive-binning method with the criterion of TS$\geq9$ for each time bin, where the minimum time bin is taken as one day. The derived flux of each time bin has a large error bar, as illustrated in Figure \ref{LC}(c). However, many serial time bins only need the minimum bin size of one day to meet TS$>$9, and the variability on a daily timescale for PKS 1413+135 can still be observed. Note that PKS 1413+135 has been announced to be detected at the VHE band with MAGIC telescopes on January 12, 2022 (MJD 59591, \citealp{2022ATel15161....1B}). This source is also in a relatively high state at the GeV band during MJD 59589--59591, as shown in Figure \ref{LC}(c). Its flux is $F_{0.1-300\rm\,GeV}\sim2.25\times10^{-10}$ erg cm$^{-2}$ s$^{-1}$ on MJD 59588, raises to $F_{0.1-300\rm\,GeV}\sim6.64\times10^{-10}$ erg cm$^{-2}$ s$^{-1}$ on MJD 59590, declines to $F_{0.1-300\rm\,GeV}\sim2.51\times10^{-10}$ erg cm$^{-2}$ s$^{-1}$ on MJD 59592, and then drops to $F_{0.1-300\rm\,GeV}\sim3.0\times10^{-11}$ erg cm$^{-2}$ s$^{-1}$ after MJD 59593. We can also observe that there are several time bins with $F_{0.1-300\rm\,GeV}>10^{-9}$ erg cm$^{-2}$ s$^{-1}$ between MJD 58700--59000, indicating that PKS 1413+135 is in a high-flux state at the GeV band and there may be detectable VHE $\gamma$-rays during that time.

In the X-ray band, there are 3 \emph{XMM-Newton}, 6 \emph{Chandra}, and 16 \emph{Swift}-XRT observations from March 2007 to January 2022. As shown in Figure \ref{LC}(a), most of them cluster at the time after January 2020. Considering the large errors of data points observed by the \emph{Swift}-XRT, a weighted mean flux $\langle{F}\rangle$ and its variance $\sigma_{\langle{F}\rangle}$ are estimated by (\citealp{1996ApJ...473..763M})
\begin{equation}
\langle{F}\rangle=\left[\sum\limits_{i=1}^N\frac{F_i}{\sigma^{2}_i}\right]\left[\sum\limits_{i=1}^N\frac{1}{\sigma^{2}_i}\right]^{-1},
\end{equation}
\begin{equation}
\sigma^{2}_{\langle{F}\rangle}=\left[\sum\limits_{i=1}^N\frac{1}{\sigma^{2}_i}\right]^{-1},
\end{equation}
where $N$ is the number of the data points, ${F_i}$ and $\sigma_{i}$ are the flux and its error for the $i$th data point. We obtain $\langle{F}\rangle_{2-10 \rm\,keV}=(6.97\pm0.17)\times10^{-13}$ erg cm$^{-2}$ s$^{-1}$ for the 25 observational data points, which is shown as the horizontal dashed line in Figure \ref{LC}(a). Interestingly, the last two points in Figure \ref{LC}(a) are obtained with the observations on January 13--14 2022, just after the ATel of the VHE detection for PKS 1413+135. We find that the X-ray flux declines from $1.3\times10^{-12}$ erg cm$^{-2}$ s$^{-1}$ on January 13, 2022 (MJD 59592) to $5.9\times10^{-13}$ erg cm$^{-2}$ s$^{-1}$ on January 14, 2022 (MJD 59593), a factor of $\sim2$ within one day. \cite{2002AJ....124.2401P} reported that the ASCA observations yield a flux of $F_{2-10\rm\,keV}=9\times10^{-13}$ erg cm$^{-2}$ s$^{-1}$, which declined a factor of $\sim$5 compared to the one seen in previous X-ray observations, but the previous X-ray flux obtained from the \emph{ROSAT} and \emph{Einstein} observations is almost completely based on the extrapolations. These results indicate that the X-rays of PKS 1413+135 are also variable.

\begin{figure}
 \centering
  \includegraphics[angle=0,scale=1.]{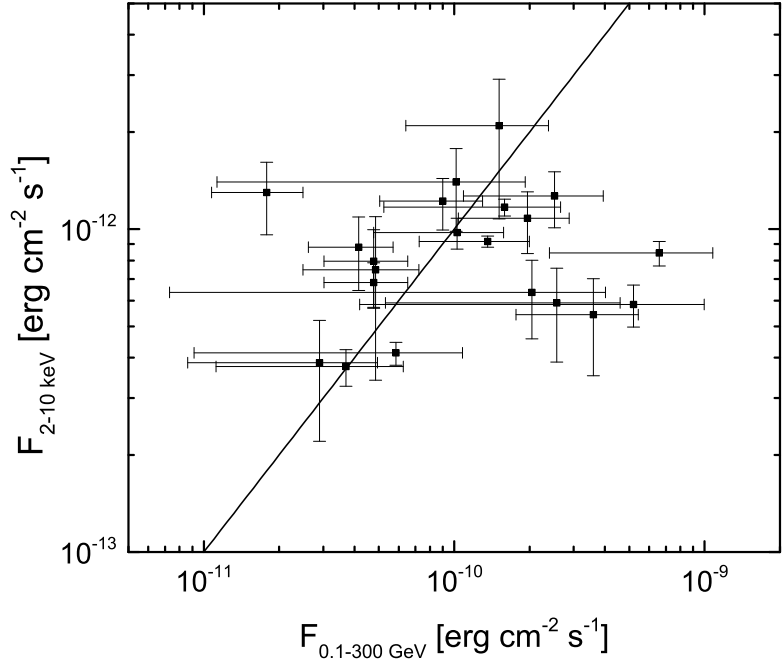}
\caption{Correlation of flux between $\gamma$-rays and X-rays for PKS 1413+135. The solid line represents $F_{2-10\rm\,keV}=F_{0.1-300\rm\,GeV}/100$. The data are taken from Figure \ref{LC}.}\label{G-X}
\end{figure}

To quantify the variability of X-rays, we calculate $\chi^2=\sum\limits_{i=1}^N\frac{(F_i-\langle{F}\rangle)^2}{\sigma_{i}^2}$ and the associated probability $p(\chi^2)=1-p(>\chi^2)$ (\citealp{2022MNRAS.515.1723C} and references therein) of the light curve. We find that $\chi^2=245.9$ (the number of the degrees of freedom d.o.f.$=N-1=24$) with $p\ll10^{-7}$, indicating that the variability of X-rays is also far beyond $5\sigma$ confidence level. We also calculate the fractional variability amplitude $F_{\rm var}$ and its uncertainty for the X-ray light curve via (\citealp{2022MNRAS.515.1723C} and references therein)
\begin{equation}
F_{\rm var}=\frac{\sqrt{\sum\limits_{i=1}^N[(F_i-\langle{F}\rangle)^2-\sigma^2_{i}]/N}}{\langle{F}\rangle},
\end{equation}
\begin{equation}
\sigma_{\rm var}=\sqrt{\left(\sqrt{\frac{\sigma^2_{\langle{F}\rangle}}{N}}\cdot\frac{1}{\langle{F}\rangle}\right)^2+\left(\sqrt{\frac{1}{2N}}\cdot\frac{\sigma^2_{\langle{F}\rangle}}{\langle{F}\rangle^2F_{\rm var}}\right)^2},
\end{equation}
where $\langle{F}\rangle$ and $\sigma_{\langle{F}\rangle}$ are the mean flux and its variance derived with Equations (2) and (3). $F_{\rm var}=0.456$ and $\sigma_{\rm var}=0.005$ are yielded for the X-ray light curve in Figure \ref{LC}(a). Further more, we estimate the fractional variability amplitude with Equations (2)--(5) for the $\gamma$-ray light curves in Figures \ref{LC}(b) and \ref{LC}(c), and obtain $F_{\rm var}=14.990\pm0.005$ for the whole $\gamma$-ray light curve and $F_{\rm var}=7.472\pm0.003$ for the $\gamma$-ray light curve after MJD 58500, respectively. Clearly, the variability amplitude of $\gamma$-rays is much larger than that of X-rays. We also investigate the correlation of variability between X-rays and $\gamma$-rays using the observation data in Figures \ref{LC}(a) and \ref{LC}(c), as displayed in Figure \ref{G-X}. Considering the large errors of data points, we use a bootstrap method to estimate the correlation coefficient ($r$) between $F_{2-10\rm\,keV}$ and $F_{0.1-300\rm\,GeV}$, and obtain $r\sim0.1$. Hence, no correlation of flux variation between two bands is observed.

In the radio band, the significant variations have been widely reported and studied by the community (e.g., \citealp{1996AJ....111.1839P, 2017ApJ...845...89V, 2021ApJ...907...61R, 2022ApJ...927...24P}). In the optical-UV band, the intrinsic optical-UV emission of the source is affected by the foreground edge-on spiral galaxy and is strongly extinct (\citealp{1992ApJ...400L..17S, 2002AJ....124.2401P, 2021ApJ...907...61R}). This source is highly variable in radio and $\gamma$-ray bands, hence its intrinsic optical emission may be potentially variable.

\section{Spectral Variation}

\begin{figure}
 \centering
  \includegraphics[angle=0,scale=0.35]{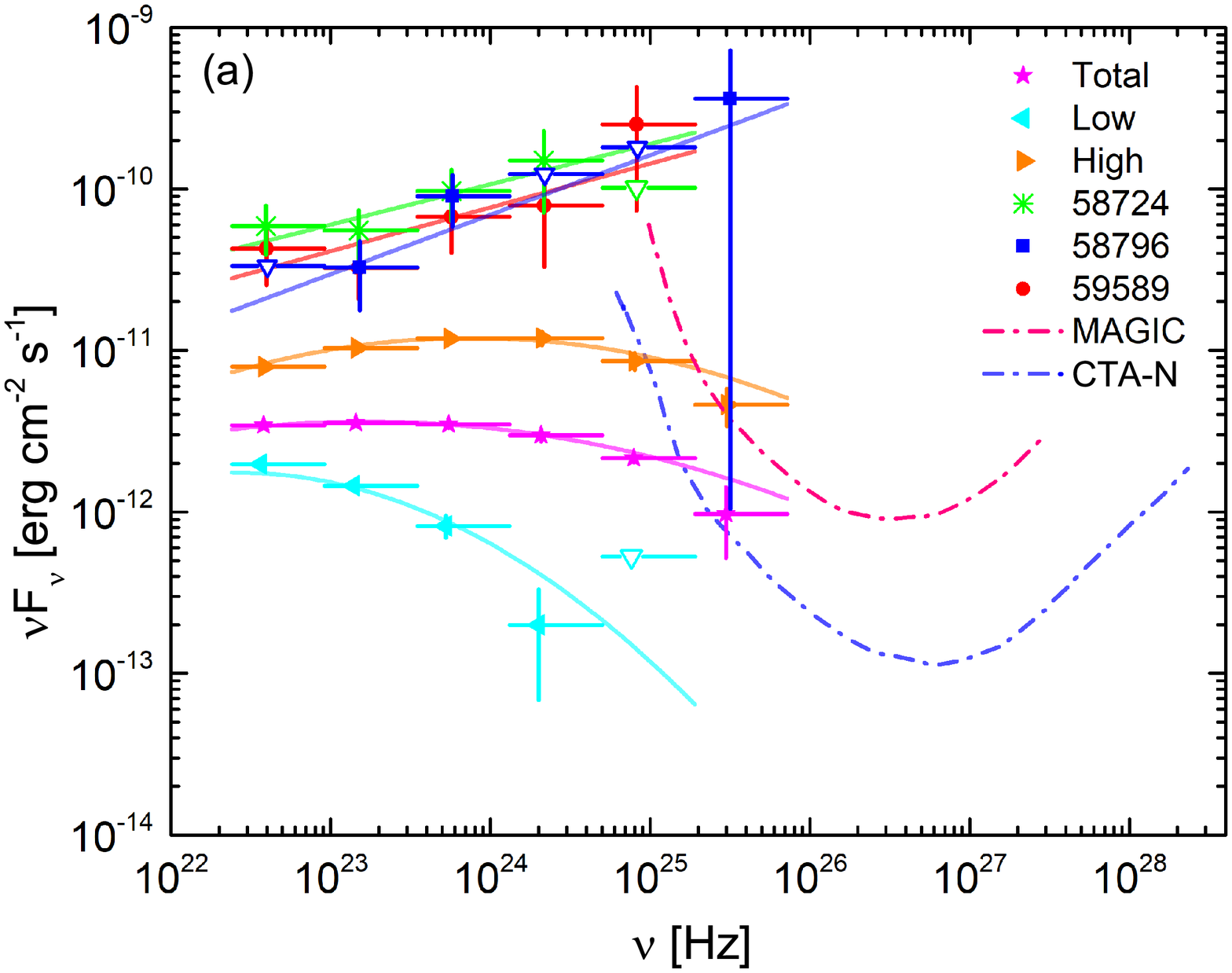}
  \includegraphics[angle=0,scale=0.7]{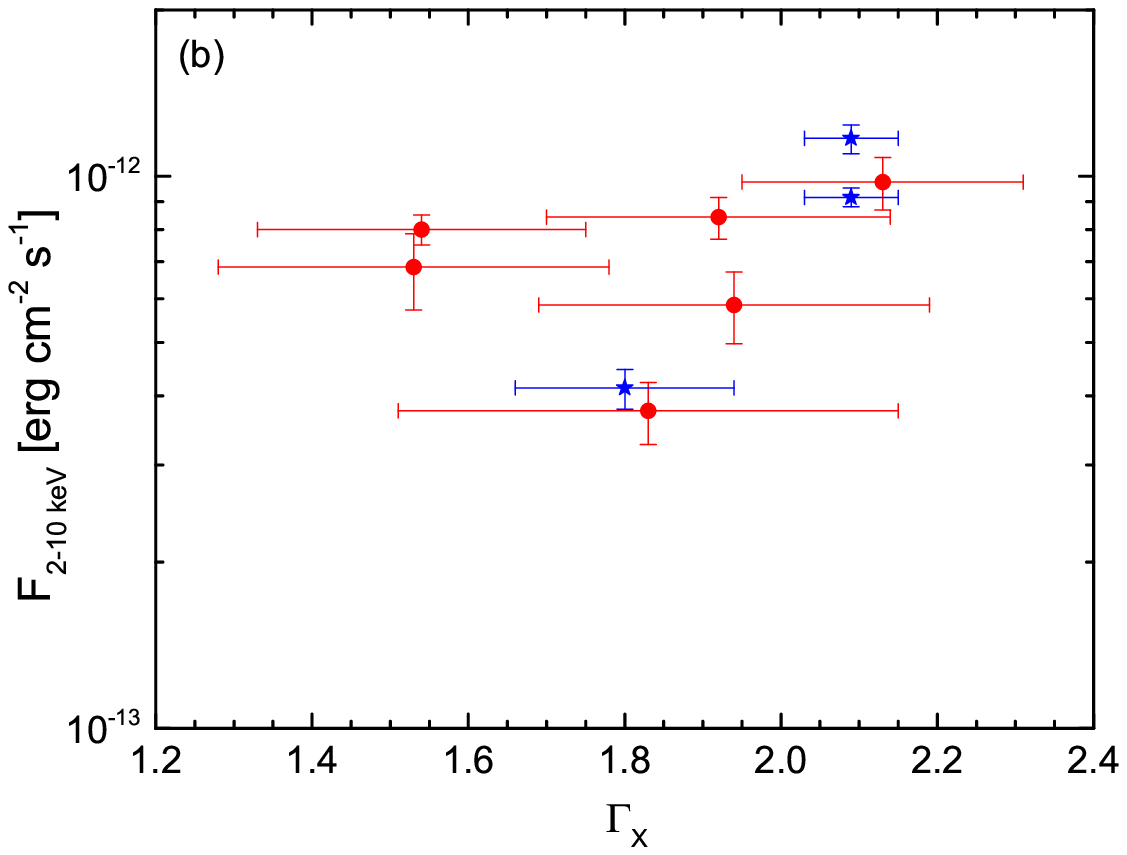}
\caption{\emph{Panel-(a)}: the spectra observed by the \emph{Fermi}-LAT for PKS 1413+135, including the time-integrated spectrum of the whole light curve (magenta symbols) in Figure \ref{LC} together with the time-integrated spectra before (MJD 54682--58500, cyan symbols) and after (MJD 58500--59660, orange symbols) MJD 58500, the time-resolved spectra of MJD 58724-58726 (green symbols), MJD 58796--58798 (blue symbols), and MJD 59589--59591 (red symbols). The corresponding colored solid lines represent the fitting results. The sensitivity curves of MAGIC telescopes (pink dot-dashed line, 50 hr) and CTA-N (blue dot-dashed line, 50 hr) are also given in the figure. \emph{Panel-(b)}: $\Gamma_{\rm X}$ against $F_{2-10\rm\,keV}$, where $\Gamma_{\rm X}$ and $F_{2-10\rm\,keV}$ are the photon spectral index and flux in the 2--10 keV band. They are obtained with the observations of \emph{Chandra} (red circles) and \emph{XMM-Newton} (blue stars).}\label{spectra}
\end{figure}

The time-integrated spectrum of the $\sim$13.5-year \emph{Fermi}-LAT observations derived from the likelihood fit is well described with the log-parabola function, as shown in Figure \ref{spectra}(a). The fitting parameters are reported in Table 1. As mentioned above, the $\gamma$-ray emission of PKS 1413+135 can be divided into two stages with a division of MJD 58500. We also extract the time-integrated spectra in the two stages. It is found that the time-integrated spectra of the two stages are also well fitted by the log-parabola function and a smaller curvature parameter value with a harder spectral index is presented in the high-flux stage than in the low-flux stage. The results are also given in Figure \ref{spectra}(a) and Table 1. One can observe that the spectrum in the high-flux stage is harder than that in the low-flux stage.

To further reveal the spectral variation feature, we make time-resolved spectral analysis for the observational data in the following time slices, MJD 58724--58726, MJD 58796--58798, and MJD 59589--59591, namely, the source being in the high-flux stage. The time slice of MJD 58724--58726 is within the time bin of the highest-flux point in Figure \ref{LC}(b). The VHE observation with MAGIC telescopes is conducted in the time slice of MJD 59589--59591. In the time slice of MJD 58796--58798, \emph{Fermi}-LAT detects the maximum energy photon ($\sim$236 GeV) from the source in its $\sim$13.5-year observations. The spectral analysis results are given in Figure \ref{spectra}(a) and Table 1. As illustrated in Figure \ref{spectra}(a), the obvious spectral evolution at the GeV band is presented for PKS 1413+135. Firstly, the flux variation in the high energy end is more prominent than that in the low energy end. Above $10^{24}$ Hz, the flux variation is almost three orders of magnitude. Secondly, the three time-integrated spectra need a curved log-parabola function to fit with a softer photon spectral index, while the three time-resolved spectra of $\gamma$-ray flares can be well modelled by a simple power-law function with a harder photon spectral index. For the power-law function, we can regard it as a log-parabola function with the curvature parameter $\beta=0$. As listed in Table 1 and shown in Figure \ref{spectra}(a), the spectral curvature decreases and the spectrum becomes hard along with the increase of flux, indicating a ``harder when brighter'' behavior in the GeV band for PKS 1413+135.

In the X-ray band, we generate the \emph{Chandra} and \emph{XMM-Newton} spectra at different epochs as described in Section 2.2, and obtain the values of $F_{2-10\rm\,keV}$ and $\Gamma_{\rm x}$ for nine observational epochs. We show $F_{2-10\rm\,keV}$ against $\Gamma_{\rm x}$ in the $F_{2-10\rm\,keV}-\Gamma_{\rm x}$ plane, as displayed in Figure \ref{spectra}(b). Considering the small sample statistics and large errors of data points, we also use the bootstrap method to estimate the correlation coefficient ($r$) of the $F_{2-10\rm\,keV}-\Gamma_{\rm x}$ relation and obtain $r=0.35$, and thus no obvious correlation between $F_{2-10\rm\,keV}$ and $\Gamma_{\rm x}$ is found for PKS 1413+135.

\section{$\gamma$-ray Emission Property Derived from Broadband SED Modeling}

As mentioned above, the GeV $\gamma$-ray emission has a ``harder when brighter'' spectral variation behavior, and the maximum energy of the detected photons is up to $\sim$236 GeV, as listed in Table 1, where the maximum energy of the detected photons is estimated with the \emph{gtsrcprob} tool. We show the sensitivity curves of MAGIC telescopes (50 hours) and Cherenkov Telescope Array (CTA) North array (CTA-N, 50 hours) in Figure \ref{spectra}(a), where the sensitivity curves are obtained from \cite{2016APh....72...76A} and the CTA webpage\footnote{https://www.cta-observatory.org/science/cta-performance/}, respectively. The high energy end of the time-resolved spectrum during MJD 59589--59591 is over the sensitivity of MAGIC telescopes and the hard spectrum indicates that the GeV spectrum would extend to the VHE band. Interestingly, PKS 1413+135 was announced to be detected by the MAGIC telescopes on January 12, 2022 (MJD 59591, \cite{2022ATel15161....1B}). The high flux with a very hard spectrum ($\Gamma_{\gamma}=1.63\pm0.14$) between MJD 58796--58798, especially the detection of the maximum energy photon at $\sim$236 GeV, clearly implies that the flux of PKS 1413+135 at the VHE band should be detectable by the MAGIC telescopes during that time. As given in Table 1, the maximum energy of the detected photons during MJD 59589--59591 and MJD 58724--58726 for PKS 1413+135 is 25.7 GeV and 55.7 GeV, respectively. Since only one photon with energy of $\sim$55 GeV was detected during MJD 58724--58726, an upper-limit is given for the highest energy bin in the time-resolved spectrum. Although the average flux for the time slice of MJD 58724--58726 is highest among the six time slices in Table 1, the slight soft spectrum, especially, the upper-limit for the highest energy bin, may indicate that the simultaneous VHE emission of PKS 1413+135 is below the sensitivity of MAGIC telescopes. We observe that even the time-integrated spectrum of the whole high-flux stage (MJD 58500--59660) is also marginally over the sensitivity of MAGIC telescopes, but the source may not be detectable by MAGIC telescopes at the VHE band when its GeV $\gamma$-ray flux is low.

\begin{figure}
 \centering
  \includegraphics[angle=0,scale=0.7]{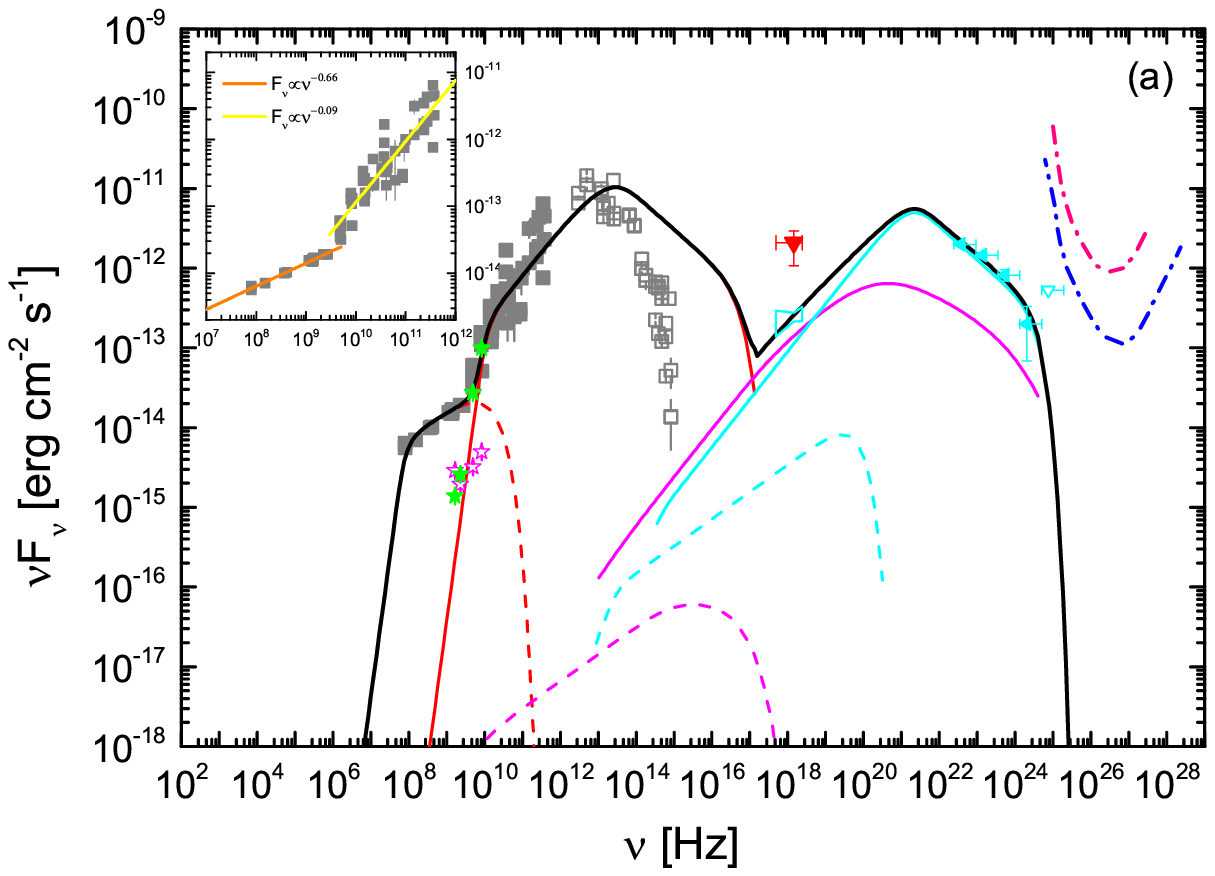}
  \includegraphics[angle=0,scale=0.7]{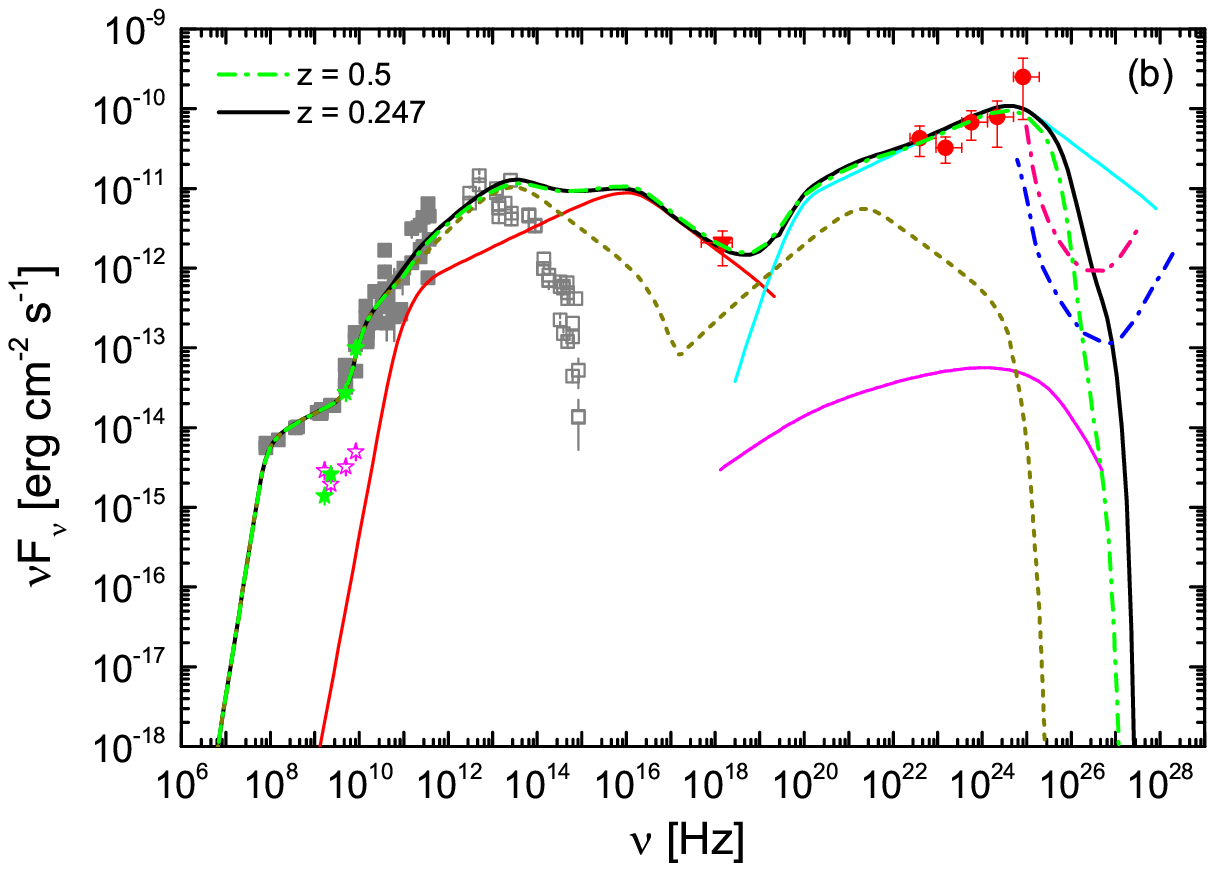}
\caption{Observed SEDs with model fitting for PKS 1413+135 in the low-flux (\emph{Panel-(a)}) and high-flux (\emph{Panel-(b)}) states of $\gamma$-rays. The $\gamma$-ray spectra in the two panels are the time-integrated spectrum before MJD 58500 (\emph{Panel-(a)}) and the time-resolved spectrum during MJD 59589--59591 (\emph{Panel-(b)}), respectively. The data in radio--optical bands, the gray solid and opened squares, are taken from the NED. The green solid and magenta opened stars indicate the data of the radio core and knot-D, which are taken from Perlman et al. (1996). In the X-ray band, the red inverted triangle represents the highest flux point in Figure \ref{LC}(a) observed by the \emph{Swift}-XRT on March 26, 2020 (MJD 58934), which is the integral flux value of $F_{2-10\rm\,keV}$. The cyan bowtie is obtained with the lowest flux point in Figure \ref{LC}(a) observed by \emph{Chandra} on December 20, 2019 (MJD 58837), which indicates the flux at 6 keV with $\Gamma_{X}=1.83\pm0.32$. The sensitivity curves of MAGIC telescopes (pink dot-dashed line, 50 hr) and CTA-N (blue dot-dashed line, 50 hr) are also presented. \emph{Panel-(a)}: The black solid line is the sum of each component emission for the source located at z=0.247, synchrotron radiation (red lines), SSC process (magenta lines), and EC process (cyan lines), where the solid and dashed lines respectively represent the radiations from the sub-pc-/pc-scale jet component and the large-scale jet extended region. An insert in the upper left of Panel-(a) shows the fitting results for the spectra from 80 MHz to 2.7 GHz and from 4.8 GHz to 375 GHz with the power-law function of $F_{\nu}\propto\nu^{-\alpha}$, and $\alpha\sim0.66$ (orange solid line) and $\alpha\sim0.09$ (yellow solid line) are obtained, respectively. \emph{Panel-(b)}: The dark-yellow dashed line is the same line as the black solid line in \emph{Panel-(a)}. The colored solid lines represent the different radiation components of the compact $\gamma$-ray emission region, synchrotron radiation (red line), SSC process (magenta line), and EC process (cyan line), where the cyan line at the VHE band indicates the intrinsic flux of the source without considering the EBL absorption. The black solid line is the sum of the dark-yellow dashed line and three colored solid lines after considering the EBL absorption for the source located at z=0.247. The green dot-dashed line represents the fitting result when the source is located at $z = 0.5$.}\label{SED}
\end{figure}

To further study the $\gamma$-ray emission property of PKS 1413+135, we collect the data at low-energy band from the NASA/IPAC Extragalactic Database (NED, \citealp{NASA/IPAC Extragalactic Database (NED). 2019}), together with the X-ray and $\gamma$-ray data derived in this paper, and construct its broadband SEDs in the low and high states of $\gamma$-ray emission, respectively, as illustrated in Figure \ref{SED}. The \emph{Fermi}-LAT time-resolved spectrum of MJD 59589--59591, corresponding to the first detection of the source at VHE band, is considered as the data of high-flux state, and the highest flux point in Figure\ref{LC}(a) observed by the \emph{Swift}-XRT on March 26, 2020 (MJD 58934) is taken as the upper-limit of X-rays. The \emph{Fermi}-LAT time-integrated spectrum of MJD 54682--58500 and the \emph{Chandra} observation data on December 20, 2019 (MJD 58837)\footnote{It is the data point with the lowest flux in Figure\ref{LC}(a) and also corresponds to a low-flux state at GeV band.} are set as the low-state data. As shown in Figure \ref{SED}, it seems that there are two different components at the radio band, similar feature being observed in some blazars (e.g., \citealp{2015MNRAS.452.3457G}). Using a power-law function of $F_{\nu}\propto\nu^{-\alpha}$ to fit them, we gain $\alpha=0.66\pm0.01$ from 80 MHz to 2.7 GHz and $\alpha=0.09\pm0.06$ from 4.8 GHz to 375 GHz, respectively; the steep radio spectrum below a few GHz, $\alpha\sim0.66$, indicates the signature of the optically thin synchrotron emission from the large-scale jet extended regions. the flat radio spectrum from a few GHz up to $\sim10^3$ GHz, $\alpha\sim0.09$, is produced by the superposition of the emission from several jet structures located at sub-pc and pc scale. The $\gamma$-ray emission region is too compact to produce the observed radio radiations below $\sim10^3$ GHz on account of the synchrotron-self-absorption (SSA) effect (e.g., \citealp{2015MNRAS.452.3457G}).

If the $\gamma$-ray emission region is inside of the broad line region (BLR), there would be the absorption (\citealp{2006ApJ...653.1089L}) and Klein--Nishina (KN, \citealp{2009MNRAS.397..985G}) effects. Especially the detection of the high-energy photon at 236 GeV by the \emph{Fermi}-LAT, we thus propose that the $\gamma$-ray emission region is outside of the BLR. And the torus would provide the seed photons of the external Compton (EC) process. The synchrotron radiation energy density can be estimated by $U^{'}_{\rm syn}=\frac{L_{\rm syn}}{4\pi R^2c\delta^4}$ (\citealp{1998ApJ...509..608T}), where $L_{\rm syn}=(\frac{1}{1-\alpha_1}+\frac{1}{\alpha_2-1})L_{\rm s}$, $L_{\rm s}$ is taken as the luminosity at $\sim5\times10^{12}$ Hz (z=0.247), $\alpha_1=0.73$ and $\alpha_2=1.34$ (the spectral indices at the GeV band in Figure \ref{SED}(a) and (b), respectively), the Doppler boosting factor $\delta\sim10$ (\citealp{2022ApJ...927...24P}). We obtain $U^{'}_{\rm syn}\sim4.9\times10^{-7}$ erg cm$^{-3}$ for $R=1$ pc and $U^{'}_{\rm syn}\sim4.9\times10^{-3}$ erg cm$^{-3}$ for $R=0.01$ pc, respectively. The energy density of torus in the comoving frame is $U^{'}_{\rm IR}\sim3\times10^{-4}\Gamma^2$ erg cm$^{-3}$ (e.g., \citealp{2009MNRAS.397..985G, 2014ApJS..215....5K}), where $\Gamma$ is the bulk Lorenz factor of the emission region. Hence, the contribution of the EC/torus process comparing with the synchrotron-self-Compton (SSC) component should be taken into account if the emission region is located at sub-pc or pc scale from the core.

The electron distribution in radiation region is taken as a power-law or broken power-law. It is characterized by an electron density parameter ($N_0$), a break energy $\gamma_{\rm b}$, and indices ($p_1$ and $p_2$) in the range of $\gamma_{\rm e}$ to $[\gamma_{\min}, \gamma_{\max}]$, where $\gamma_{\rm e}$ is the Lorentz factor of electrons. The radiation region is assumed as a sphere with radius $R$, magnetic field strength $B$, the Doppler boosting factor $\delta$, where $\delta=1/(\Gamma-\sqrt{\Gamma^2-1}\cos\theta)$, $\Gamma$ and $\theta$ are the bulk Lorenz factor and viewing angle of the emission region. The synchrotron (syn), SSC, and EC processes of the relativistic electrons are considered to fit the broadband SEDs of PKS 1413+135\footnote{The same code was used in other $\gamma$-ray emitting compact radio sources (\citealp{2020ApJ...899....2Z, 2021RAA....21..201G, 2022ApJ...927..221G}).}. The spectrum from near-infrared to optical-UV bands is seriously extinct due to an edge-on intervening Seyfert 2 galaxy between PKS 1413+135 and the Earth (\citealp{1992ApJ...400L..17S, 1994ApJ...424L..69P}), and thus we would not consider these data during the SED modeling. The KN effect and the absorption of high-energy $\gamma$-ray photons by extragalactic background light (EBL, \citealp{2008A&A...487..837F}) are taken into account during the SED modeling.

The radio spectrum below several GHz in the broadband SED of PKS 1413+135 is thought to be produced by the emission of jet extended regions. The overall size of PKS 1413+135 at radio band is $\sim$110 mas (\citealp{2021ApJ...907...61R}), which corresponds to the projection distance of $\sim$422 pc for z=0.247. As reported by \cite{2008ApJ...680..911S}, the external photon field is still dominated by the dusty torus at this scale and its energy density ($U_{\rm IR}$) can be estimated by Equation (21) in \cite{2008ApJ...680..911S}. We obtain $U_{\rm IR}\sim9\times10^{-9}$ erg cm$^{-3}$, which is much higher than the energy density of cosmic microwave background (CMB) light at z=0.247. Hence the syn+SSC+EC/torus model under the equipartition condition ($U_B=U_{\rm e}$) is taken to reproduce the emission of the large-scale jet extended region, where $U_B$ and $U_{\rm e}$ are the energy densities of the magnetic fields and electrons. $R$ is taken as $R=211$ pc, corresponding to the half of overall size at the radio band. The relativistic effect is not considered, i.e., $\delta=\Gamma=1$. Considering the limit observation data of the large-scale jet extended region at radio band, the electron distribution is taken as a power-law. $p_1$ is derived by the radio spectral index ($\alpha\sim0.66$) below several GHz and is fixed as $p_1=2.32$. $\gamma_{\min}=1$ is also fixed during the modeling. We adjust the values of $\gamma_{\max}$ and $N_0$ to fit the radio spectrum below several GHz in the broadband SED and obtain $\gamma_{\max}=2000$ and $N_0=0.036$ cm$^{-3}$. And then we fix the reproduced spectrum component of the large-scale jet extended region in the following SED modeling for the low-flux and high-flux states.

The GeV emission of the low-flux and high-flux states may have different origins, and we return to this point in Section 6.1. For the low-flux state of GeV emission, we propose that the emission is dominated by the sub-pc-/pc-scale jet structures. The radius of radiation region is taken as $R=0.6$ pc, corresponding to the half of the projection distance between the radio core and component-D8 (0.16 mas, \citealp{2022ApJ...927...24P}) for z=0.247. Considering the relativistic effect, $\delta=\Gamma$ is assumed during SED modeling, namely, the viewing angle ($\theta$) being equal to the opening angle ($1/\Gamma$) of jet (e.g., \citealp{2021ApJ...906..105C}). The electron distribution of a broken power-law is used here. $\gamma_{\min}=1$ and $\gamma_{\max}=10^5$ are fixed. $p_2=3.84$ is fixed and constrained by the spectral index in GeV band. We adjust the values of $B$, $\delta$, $p_1$,  $\gamma_{\rm b}$, and $N_0$ to fit the SED of the low-flux state in Figure \ref{SED}(a), and then obtain $B=0.45$ G, $\delta=2.2$, $p_1=1.8$, $\gamma_{\rm b}$=2609, and $N_0=2.81$ cm$^{-3}$. For the high-flux state at GeV band, we speculate that the outbursts after MJD 58500 in Figure \ref{LC}(b) are due to the restarted activity of the central engine and originate from the region more close to the black hole. The radius of radiation region is taken as $R=\delta c\Delta t/(1+z)$, where $\Delta t=1$ day is taken since the variability on a daily timescale is presented in Figure \ref{LC}(c). As shown in Figure \ref{SED}(b), the peak of the second bump is above $10^{25}$ Hz, which needs an electron population with higher energy than that produces the low-flux emission. We thus consider the syn+SSC+EC processes of another electron population to reproduce the emission of the GeV outburst on the basis of the fitting result of the low-flux state. The distribution of the new electron population is also taken as a broken power-law. $p_1=2.46$, $p_2=3.84$, and $\gamma_{\max}=10^6$ are fixed, where $p_1$ is derived by the spectral index in GeV outburst while $p_2$ is taken the same value as the low-energy electron population\footnote{Although no observational data are available to constrain the $p_2$ value, a broken power-law (not a power-law) distribution of electrons is generally needed to explain the broadband SEDs of $\gamma$-ray emitting AGNs (\citealp{1998ApJ...509..608T, 2014A&A...563A..90A, 2015MNRAS.452.3457G}), we thus still consider a broken power-law distribution for the higher-energy electron population. }. $\gamma_{\min}=50$ is roughly constrained by the upper-limit of X-ray flux. We adjust the values of $B$, $\delta$, $\gamma_{\rm b}$, and $N_0$ to fit the spectrum during $\gamma$-ray outburst on the basis of the fitting result of the low-flux state. No observational data at optical-UV band are available, and the highest flux point in Figure\ref{LC}(a) observed by the \emph{Swift}-XRT is taken as the upper-limit to roughly constrain the fitting parameters. We obtain $B=0.8$ G, $\delta=20$, $\gamma_{\rm b}=1.45\times10^4$, and $N_0=631.8$ cm$^{-3}$. The fitting results are shown in Figure \ref{SED} and the modeling parameters are listed in Table \ref{table3}. Note that $B$ and $\delta$ are degenerate (e.g., \citealp{2012ApJ...752..157Z}), and the model parameters cannot be totally constrained with the current observational data.

As displayed in Figure \ref{SED}, the $\gamma$-rays are absolutely produced by the EC process while the X-rays are a hybrid of several components. The observed spectral indices at the X-ray band cluster at $1.8<\Gamma_{\rm X}<2.2$, as shown in Figure \ref{spectra}(b), suggesting a transition between the synchrotron radiation and inverse Compton component (e.g., \citealp{1998MNRAS.301..451G}). The complex origins of X-rays may result in the smaller amplitude of flux variation than that of $\gamma$-rays. No significant correlation of variability between X-rays and $\gamma$-rays, as shown in Figure \ref{G-X}, also demonstrates that the X-rays have the different origin from $\gamma$-rays.

The observation time of the $\gamma$-ray spectrum in Figure \ref{SED}(b) corresponds to the time when PKS 1413+135 was announced to be detected firstly at the VHE band by the MAGIC telescopes. The predicted flux by the SED modeling at the VHE band exceeds the sensitivity of MAGIC telescopes, whether considering the EBL absorption or not. The integral flux of the model prediction over the MAGIC sensitivity is $\sim1.4\times10^{-10}$ erg cm$^{-2}$ s$^{-1}$ after considering the EBL absorption. Recently, \cite{2022MNRAS.515.4505M} reported that the attenuation of VHE photons due to the EBL absorption may be overestimated, especially for the high redshift sources. Hence, there may be a higher observational VHE flux for PKS 1413+135 than the model prediction. We can also observe that PKS 1413+135 would not be detectable at the VHE band when its GeV emission is in the low-flux state, as illustrated in Figure \ref{SED}(a).

The host galaxy and redshift of PKS 1413+135 are also still debated. PKS 1413+135 may be a background source of the spiral galaxy and is located at a higher redshift of $0.247<z<0.5$ (\citealp{2021ApJ...907...61R}). As displayed in Figure \ref{SED}(b), the predicted flux by the SED modeling after considering the EBL absorption is still over the sensitivity of MAGIC telescopes if PKS 1413+135 is located at z=0.5. Hence, we would not give a more accurate constraint on its redshift with the SED modeling.

\section{Discussion}

\subsection{Origin of $\gamma$-ray Emission for PKS 1413+135 }

As displayed in Figure \ref{LC}(b), the GeV $\gamma$-rays of PKS 1413+135 are in a low-flux stage before MJD 58500, then undergo the obvious outbursts, and always stay in a high-flux stage after MJD 58500. We can also observe that the GeV spectral features are quite different for the high-flux and low-flux states, as displayed in Figure \ref{spectra}(a). The curvature log-parabola energy spectrum with a softer spectral index is shown in the low-flux state while the spectra in the outbursts of $\gamma$-rays can be well fitted by a power-law function with a harder spectral index. The different spectral features in the two states for PKS 1413+135 may reflect the differently dominant acceleration mechanisms or acceleration regions of relativistic electrons. In the radio band, the emission above 5 GHz is totally dominated by the radio core (Table 2 in \citealp{1996AJ....111.1839P}; Figure 4 in \citealp{2022ApJ...927...24P}). Except the radio core and knot-D, the other jet structures show the steep spectra, which should originate from the optical thin synchrotron radiations, as illustrated by the red dashed line in Figure \ref{SED}(a). The inverted spectrum of radio core is due to the SSA effect while the flat spectrum ($\alpha\sim0$) of knot-D may indicate a reacceleration and/or recollimation region (\citealp{1996AJ....111.1839P}). Component-D8 is the closest component to the radio core located at 0.32 mas away (\citealp{2022ApJ...927...24P}) and is also the brightest component except radio core in the Monitoring Of Jets in Active galactic nuclei with VLBA Experiment (MOJAVE) images (\citealp{2019ApJ...874...43L}). We speculate that the reacceleration spectral feature of knot-D is due to the emergence of a new component-D8 from the radio core, and the component-D8 interacts with other jet structures to reaccelerate the electrons in knot-D. \cite{2022ApJ...927...24P} reported that component-D8 displayed the rapid variability between 1995 and 1998. The long-term light curve of PKS 1413+135 at 14.5 GHz taken from the University of Michigan Radio Astronomical Observatory (UMRAO, \citealp{2021ApJ...907...61R}) shows three outbursts around in 1982, 1988, and 1992. Maybe the emerging of component-D8 is connected with one of the three outbursts, and this is roughly coincident with the derived result using the separation speed (9.2$\pm$2.2 $\mu$as/yr, \citealp{2019ApJ...874...43L}) and distance of component-D8 from the radio core. We thus suggest that the $\gamma$-rays of PKS 1413+135 may have two origins: one is from the sub-pc-/pc-scale jet components, which contributes the low-flux $\gamma$-rays in Figure \ref{LC}(b) and the flat radio spectral component in Figure \ref{SED}(a). another is connected with the outbursts of $\gamma$-rays, which may be due to the ejection of a new component recently from the core, similar to many blazars (e.g., \citealp{2010ApJ...710L.126M, 2013ApJ...773..147J, 2017MNRAS.468.4478L, 2018ApJ...858...80R, 2019MNRAS.486.2412L, 2020ApJ...899....2Z}), namely, the restarted nuclear jet activity of PKS 1413+135.

The $\gamma$-ray luminosity of source during outbursts is up to $\sim10^{47}$ erg s$^{-1}$ ($z=0.247$), as listed in Table 1, together with the significant variability at the $\gamma$-ray band, implying the strong Doppler boosting effect of the $\gamma$-ray emission region. We assume $\delta=\Gamma$ during the SED fitting, the viewing angle ($\theta$) is equal to the opening angle ($1/\Gamma$) of jet (e.g., \citealp{2021ApJ...906..105C}), and thus it is $\theta\sim3^\circ$. This is conflicted with the two-side pc-scale structure in the radio morphology of PKS 1413+135 since the symmetric radio structure of CSOs is thought to be due to a misaligned jet to the observers (\citealp{1980ApJ...236...89P, 1994ApJ...432L..87W, 1996ApJ...460..612R}). Recently, the episodic nuclear jet activity in $\gamma$-ray emitting CSOs, TXS 0128+554 (\citealp{2020ApJ...899..141L}) and CTD 135 (\citealp{2021RAA....21..201G}), was reported. We thus propose that there are episodic nuclear jet activities in the center of PKS 1413+135. The axis of the recently restarted jet is aligned within a few degree to the line of sight and is different from the direction of the pre-existing components, similar to the typical radio galaxy 3C 84 (\citealp{2016AN....337...69N}). The restarted jet may result in a new component ejected from the core, which would interact with the surrounding materials (or the pre-existing components) and then accelerate particles to produce the significant $\gamma$-ray outburst and variability. One of the most striking evidence of episodic nuclear jet activity is that two or more pairs of distinct radio components are observed on opposite sides of the radio core (\citealp{2021A&ARv..29....3O}). The outer components of jet and counter-jet for PKS 1413+135, which are far away from the radio core, such as the components A, F and G (\citealp{1996AJ....111.1839P}), may be the remnants of the nuclear jet activity long-ago.

The overall size of the radio morphology for PKS 1413+135 is $\sim$422 pc (projection size, $z=0.247$, \citealp{2021ApJ...907...61R}), and an explanation of the size below 1 kpc for compact radio sources is that they are transient or episodic sources (\citealp{2021A&ARv..29....3O} for a review), which cannot provide enough energy for the jet to propagate to large scale. On the other hand, a fairly dense nuclear medium (\citealp{1996AJ....111.1839P}) may also frustrate the jet propagation of PKS 1413+135. And the interaction between the jet and surrounding medium may change the jet direction and result in the bent in the extended jet structures. As given in Table \ref{table3}, the derived values of $\delta$ by the SED modeling are 2.2 and 20 for low-flux and high-flux states, respectively, implying that the jet of PKS 1413+135 is decelerated effectively on sub-pc scale. If the scenario described above is true, the recent $\gamma$-ray outbursts of PKS 1413+135 should be connected with the restarted nuclear jet activity. And the ejected new component from the central engine would be resolved out with the VLBI observations in the future.

\subsection{PKS 1413+135: a CSO or a Blazar}

PKS 1413+135 was classified as either a BL Lac or a ``red quasar'' on account of the inverted and rapidly variable radio spectrum, the absence of optical emission lines, its extremely steep near-IR slope, and an IR K-band polarization of $(16\pm3)\%$ (\citealp{1981Natur.293..714B, 1992ApJ...400L..17S}). We find that PKS 1413+135 also displays the significant flux variation in the GeV band accompanied by the obvious spectral evolution. As shown in Figure \ref{LC}(b), the variation of $\gamma$-ray flux for PKS 1413+135 exceeds two orders of magnitude, indicating the typical relativistic jet effect, similar to blazars. Especially, the report of the VHE detection for PKS 1413+135 by the MAGIC telescopes further supports that it may be a BL Lac since most of the confirmed extragalactic VHE emission sources are BL Lacs\footnote{http://tevcat.uchicago.edu/}. Generally, the broadband SEDs of BL Lacs can be reproduced with the single-zone syn+SSC model (e.g., \citealp{2012ApJ...752..157Z, 2014MNRAS.439.2933Y, 2022MNRAS.tmp.1874L}). However, this simple model is not suitable for PKS 1413+135 since the broad second bump in its broadband SED during $\gamma$-ray outbursts.

Different from blazars, no significant superluminal motions on pc-scale jet of PKS 1413+135 were reported. The maximum value is $1.72\pm0.11\ c$, i.e., 110$\pm$7 $\mu$as/yr (\citealp{2019ApJ...874...43L}), which corresponds to the component-D3 (in Figure 2 of \citealp{2022ApJ...927...24P}) located at $\sim$7 mas away from the radio core. On account of the miniature radio morphology with two-side pc-scale structure, PKS 1413+135 was suggested to be a CSO (\citealp{1996AJ....111.1839P}). And the counter-jet of PKS 1413+135 is also a strong radio source (\citealp{1996AJ....111.1839P, 2022ApJ...927...24P}). In contrast, blazars are believed to be on-axially observed to their jets and an asymmetric one-side aligned jet is observed. As shown in Figure \ref{G-X}, on average, the X-ray fluxes are lower than the $\gamma$-ray fluxes two orders of magnitude for PKS 1413+135. This feature is also not like that of the TeV emitting BL Lacs, as displayed by the broadband SEDs of TeV-selected BL Lacs in \cite{2012ApJ...752..157Z}. Hence, the AGN type of PKS 1413+135 is still debated.

\section{Summary}

The miniature radio morphology with two-side pc-scale structure makes PKS 1413+135 to be classified as a CSO. Interestingly, it has been detected by MAGIC telescopes and is the first CSO detected at VHE band. In this paper, we comprehensively analyzed the 13.5-year \emph{Fermi}-LAT observation data of PKS 1413+135, together with its archive X-ray data observed with \emph{Swift}-XRT, \emph{Chandra}, and \emph{XMM-Newton}, to investigate its high energy radiation properties. The significant variability is presented in the long-term GeV $\gamma$-ray light curve with a confidence level far beyond 5$\sigma$. And the whole $\gamma$-ray light curve shows two distinct stages, a low-flux stage before MJD 58500 and a high-flux stage with violent outbursts after MJD 58500. The clear spectral variation accompanying the variation of $\gamma$-ray flux is observed, namely, the curvature log-parabola energy spectrum with a softer spectral index in the low-flux state and the power-law spectrum with a harder spectral index in the $\gamma$-ray outbursts, indicating a ``harder when brighter'' behavior at GeV band. The maximum energy of the detected photons by the \emph{Fermi}-LAT during its $\sim$13.5-year observations is up to $\sim$236 GeV. The X-rays of PKS 1413+135 are also obviously variable, but no significant spectral variation is found. The variability amplitude of X-rays is much smaller than that of $\gamma$-rays, and no correlation of flux variation between two bands is presented. Attributing the broadband SEDs of PKS 1413+135 to the radiations from the compact radiation region close to the black hole, the sub-pc-/pc-scale jet component, and the large-scale jet extended region, we represented the SED by a multi-zone leptonic jet model. Its $\gamma$-rays are contributed by the EC process of relativistic electrons while the X-rays are synthesis of several radiation components. The different radiation origins may result in the small variability amplitude of X-ray flux and no correlation of variability between X-rays and $\gamma$-rays. The predicted flux by the SED modeling at the VHE band would be detectable by MAGIC telescopes when PKS 1413+135 is in a high-flux state at the GeV band. Considering the features of the temporal and spectral variations at the GeV band, together with the observations in other bands, we proposed that the $\gamma$-ray outbursts after MJD 58500 of PKS 1413+135 may be connected with the recently restarted nuclear jet activity of the central engine and are produced in a aligned core-jet component. The sub-pc and pc scale jet components, which may be produced by the previous nuclear jet activities, contribute the low-flux $\gamma$-ray emission. This scenario can be checked by the future VLBI observations.

\acknowledgments

We thank the anonymous referee for valuable suggestions. This work is supported by the National Natural Science Foundation of China (grants 12022305, 11973050, and 12133003), and Guangxi Science Foundation (grants 2018GXNSFGA281007 and 2019AC20334).

\clearpage

\clearpage

\begin{deluxetable}{lcccccccc}
\tabletypesize{\scriptsize} \tablecolumns{8} \tablewidth{0pc}
\tablecaption{\emph{Fermi}-LAT Analysis Results for PKS 1413+135}\tablenum{1}
\tablehead{\colhead{Obs-date} & \colhead{Model\tablenotemark{\scriptsize{a}}} & \colhead{$\Gamma_{\gamma}$} & \colhead{$\beta$} & \colhead{$F_{\gamma}$} & \colhead{$L_{\gamma}\tablenotemark{\scriptsize{b}}$} & \colhead{TS} & \colhead{$E_{\rm max}$} & \colhead{Det-time\tablenotemark{\scriptsize{c}}}\\
\colhead{(MJD)} & \colhead{} & \colhead{} & \colhead{} & \colhead{(erg cm$^{-2}$ s$^{-1}$)} & \colhead{(erg s$^{-1}$)} & \colhead{} & \colhead{(GeV)} & \colhead{(MJD)}}
\startdata
54682--59660&LP&$2.02\pm0.03$&$0.03\pm0.01$&$(2.28\pm0.13)\times 10^{-11}$&$(4.13\pm0.23)\times 10^{45}$&3355&235.9&58798\\
54682--58500&LP&$2.34\pm0.08$&$0.08\pm0.05$&$(5.74\pm0.61)\times 10^{-12}$&$(1.04\pm0.11)\times 10^{45}$&300&20.6&57959\\
58500--59660&LP&$1.90\pm0.03$&$0.04\pm0.01$&$(7.74\pm0.48)\times 10^{-11}$&$(1.41\pm0.09)\times 10^{46}$&4704&235.9&58798\\
58724--58726&PL&$1.75\pm0.11$&0.00 (fixed)&$(1.07\pm0.39)\times 10^{-09}$&$(1.94\pm0.71)\times 10^{47}$&200&55.7&58726\\
58796--58798&PL&$1.63\pm0.14$&0.00 (fixed)&$(8.59\pm4.02)\times 10^{-10}$&$(1.56\pm0.73)\times 10^{47}$&102&235.9&58798\\
59589--59591&PL&$1.73\pm0.13$&0.00 (fixed)&$(7.95\pm3.43)\times 10^{-10}$&$(1.44\pm0.62)\times 10^{47}$&137&25.7&59590\\
54682--59063\tablenotemark{\scriptsize{d}}&LP&$2.00\pm0.04$&$0.03\pm0.01$&$(1.94\pm0.09)\times 10^{-11}$&&&&\\
\enddata
\tablenotetext{a}{``LP'' and ``PL'' represent the spectra are fitted by the Log-Parabola (LP) and Power-Law (PL) functions, respectively.}
\tablenotetext{b}{The corresponding luminosity for the source located at $z=0.247$.}
\tablenotetext{c}{The detection time of the maximum energy photon by the \emph{Fermi}-LAT.}
\tablenotetext{d}{The analysis results given in the 4FGL-DR3 (\citealp{2022ApJS..260...53A}).}
\label{table1}
\end{deluxetable}

\begin{deluxetable}{lcccc}
\tabletypesize{\scriptsize} \tablecolumns{4} \tablewidth{0pc}
\tablecaption{Public X-ray observations of PKS 1413+135}\tablenum{2}
\label{table2}
\tablehead{\colhead{Obs-date} & \colhead{} & \colhead{Mission} & \colhead{} &  \colhead{Exp. (s)}\\
\colhead{(1)} & \colhead{} & \colhead{(2)} & \colhead{} & \colhead{(3)}}
\startdata
2007-03-20&&\emph{Chandra}&&20180\\
2019-12-20&&\emph{Chandra}&&4060\\
2019-12-29&&\emph{XMM-Newton}&&13000\\
2020-02-03&&\emph{Chandra}&&4060\\
2020-02-05&&\emph{XMM-Newton}&&17000\\
2020-03-22&&\emph{Chandra}&&4060\\
2020-05-04&&\emph{Chandra}&&4060\\
2020-06-18&&\emph{Chandra}&&4060\\
2020-07-04&&\emph{XMM-Newton}&&15000\\
2007 to 2022&&\emph{Swift}&&10$^{2}$ to 10$^{4}$\\
\enddata

\end{deluxetable}

\begin{deluxetable}{lccccccccccc}
\tabletypesize{\scriptsize} \tablecolumns{12} \tablewidth{0pc}
\tablecaption{SED Fitting Parameters of PKS 1413+135}\tablenum{3}
\tablehead{\colhead{State\tablenotemark{\scriptsize{a}}}  & \colhead{$R$} & \colhead{$R_{\rm diss}\tablenotemark{\scriptsize{b}}$}&
\colhead{$B$} & \colhead{$\delta$}& \colhead{$\Gamma$} & \colhead{$\gamma_{\min}$} & \colhead{$\gamma_{\rm b}$} &\colhead{$\gamma_{\max}$}
&\colhead{$N_{0}$ } & \colhead{$p_1$} & \colhead{$p_2$}\\
\colhead{} &\colhead{}  &\colhead{}& \colhead{(mG)} & \colhead{}& \colhead{} & \colhead{} & \colhead{} &\colhead{} &\colhead{ (cm$^{-3}$)} & \colhead{} & \colhead{}}
\startdata
E &211~pc&400~pc&1.25&1&1&1&\nodata&2000&0.036&2.32&\nodata\\
L &0.6~pc&1~pc&450&2.2&2.2&1&2609&$10^5$&2.81&1.8&3.84\\
H &$4.15\times10^{16}$~cm&$4.15\times10^{17}$~cm&800&20&20&50&14513&$10^6$&631.8&2.46&3.84\\
\enddata
\tablenotetext{a}{``E'', ``L'', and ``H'' represent the parameters for the large-scale jet extended region, the sub-pc-/pc-scale jet component, and the compact radiation region close to the black hole, respectively. The parameter values correspond to the result for the source located at $z=0.247$.}
\tablenotetext{b}{$R_{\rm diss}$ is the distance of the radiation region from the black hole, which is estimated roughly. For the large-scale jet extended region, $R_{\rm diss}$ corresponds to the overall projection size of the radio morphology. For the sub-pc-/pc-scale jet component, $R_{\rm diss}$ is roughly the projection distance of the component-D8. For the compact radiation region, $R_{\rm diss}\sim10*R$ (see also \citealp{2009MNRAS.397..985G}).}
\label{table3}
\end{deluxetable}

\end{document}